\documentclass[%
 ,secnumarabic%
,amssymb, amsmath,nobibnotes, aps, pre]{revtex4}
\usepackage[T1]{fontenc}
\usepackage[cp1250]{inputenc}
\usepackage[polish,english]{babel}
\usepackage{graphicx}
\usepackage{bm}
\usepackage{color}
\usepackage{color}
                      
\begin{document}

\title{Simulations of assortative matching: complexity of  cooperation emergence process}
\author{Paweł Sobkowicz}
\email{pawelsob@poczta.onet.pl}
\date{\today}

\begin{abstract}
The paper studies the  emergence and stability of cooperative behavior in populations of agents who interact among themselves in Prisoner's Dilemma games and who are allowed to choose their  partners. The population is then subject to evolutionary pressures, based on individual payoffs from multiple encounters. A simple formula for signaling and recognition is introduced, which allows the agents to guess prospective partner strategy and to refuse the game if the predicted outcome is unfavorable. We present both algebraic formulation of the average payoffs and results of computer simulations of evolution of such society. The simulations result in surprising variety of  behavior. We discuss possible interpretations of the results as well as relationship between artificial `computer time' and real time of possible social and biological systems that the model might be applied to.
 \end{abstract}

\maketitle

\section{Formulation of the problem}

The purpose of our work is to describe a system of agents who interact pairwise  in a simple Prisoner's Dilemma (PD) games \cite{bergstrom03-1,sobkowicz-ass}. %{\color{red} Check biblio data on Bergstrom paper}
  Based on results of such encounters, successful agents are allowed to multiply, while the least successful die out. We are interested in the evolution of such society, both in final stable states and in transient phenomena. We assume that the two types of strategies in PD game, the Cooperator and the Defector, are pure strategies for each agent, that agents have no memory, and that agents may decide whether they would play the game in a given pair or not.   The agents can, moreover  advertise or camouflage their character before the actual interaction (called here a \textbf{match}) occurs. They can also recognize   the character of other agents, although not with perfect accuracy. The purpose of this signaling and recognition, important especially for Cooperators, is to avoid, as much as possible, the  matches with Defectors. As a result the matches are no longer random, but show certain amount of assortativity, i.e. tendency of agents to pair with agents of the same type.

Such assortativity proves to be a way of defense of Cooperators against the Defectors, enabling, in some circumstances, the cooperation strategy to remain present and even to dominate the society. 
In this paper we present extension of the model described in \cite{sobkowicz-ass}. We would start with introduction of algebraic model which includes improved modeling of recognition of other agents and subsequent activities. The we would move to computer simulations which allow much ore flexibility and uncover interesting global properties of the system.

\subsection{Notation and basis of the model}
Let the total number of agents, constant in time, be  $N$. The number of Cooperators, at a given moment is $N_C$, the number of Defectors $N_D$. Fraction of Cooperators is denoted by $x=N_C/N$, of Defectors by $1-x$.

Each agent (numbered $i, j$)  is characterized by  three properties: strategy type $t(i)$ (Cooperator or Defector), signaling strength $s(i)\in [0,1]$ and  capability to minimize the error in recognizing the other agents' strategy $r(i)\in (0,1]$. The larger the value of $s(i)$, the greater chance that other agents would perceive the agent in accordance with its true type. (Obviously, for Defectors this would be a disadvantage, so Defectors would invest in minimizing the signaling, which corresponds to successful camouflage). We propose the following formula for the probability of correct identification of agent's $j$ type by agent $i$:
\begin{equation}
	P(i,j) = 1 - r(i)\left( 1-s(j)\right).
\end{equation}
For $r(i)=1$ the probability is simply given by the signal strength $s(j)$. Signaling $s(i)=0$ corresponds then  to perfect camouflage, as the probability of correct recognition would be zero. Diminishing $r(i)$ decreases the erroneous identification, for $r(i)\rightarrow 0$ agent $i$ is able to correctly recognize others almost regardless of their signal strength. It is well established that such signaling, to have useful value in showing (or hiding) real strategy must be costly \cite{frank88-1,bergstrom98-1,bergstrom98-2,noble98-1, bergstrom99-1,hasson00-1,lachman01-1, lachman03-1}.
 %{\color{red}{ Check out works of: Lachmann, M. and Bergstrom, C. T. (1998) Theor. Popul. Biol. 54, 146–160.; Grafen, A. (1990) J. Theor. Biol. 144, 517–546.; Godfray, H. C. J. (1991) Nature (London) 352, 328–330.; Bergstrom, C. T and Lachmann, M. (2001) Anim. Behav. 61, 535–543. Hasson, O. 1994. Cheating signals. Journal of Theoretical Biology 167:223–238.}}

Each agent interacts with numerous other agents in each turn. The interaction may be imagined as two phase: during the first part of the encounter both agents appraise each other and take decision whether to proceed (consent to a match) or not. The decisions are based on their signals and error reduction capabilities. If both take positive decision the actual PD game ensues, with  the results depending on the true strategies.

We envisage here two basic situations. In the first, each agent initiates a specified number of encounters $k$. For each of the encounters the initiating agent and a randomly chosen partner appraise each other, if consent  is mutual match occurs, otherwise the encounter is registered as resulting in both agents staying single. We would call this approach ``\textbf{singles allowed}''. The other approach requires the agents to search around until a suitable and consenting candidate is found. Each agent must participate in $k$ matches. This ``\textbf{must match}'' approach results is different dynamics of the population.

\subsection{Simplified analytical model vs. computer simulations}
Simulating the assortative matching described above is relatively straightforward. The values of $t(i)$, $s(i)$ and $r(i)$ are ascribed to each agent. Within each iteration each of the agents attempts to interact with a specified number of others. Results of the encounters are summed and averaged. After all encounters have taken place, the payoffs of the agents are compared and a fraction of the `best performers' are allowed to breed, replacing the characteristics of the worst performers with their own. Additionally, a very small number of agents `mutate' their characteristics between iterations to randomly chosen values within the bounds set for $t(i)$, $s(i)$ and $r(i)$. 

Details of the simulations will be presented in the later part of the paper. To help in our understanding of these results we introduce here simplified analytical formalism, based on assumption that all agents in Cooperator and Defector group  have the same values of  $s(i)$ and $r(i)$, equal to averages within the group:
\begin{equation}
	\alpha = \sum_{\text{Cooperators}} s(i) / N_C,
\end{equation}
\begin{equation}
	\beta = \sum_{\text{Defectors}} s(i) / N_D,
\end{equation}
\begin{equation}
	r_c = \sum_{\text{Cooperators}} r(i) / N_C,
\end{equation}
\begin{equation}
	r_d = \sum_{\text{Defectors}} r(i) /N_D.
\end{equation}
We would also introduce values of recognizability (by Cooperators) of Cooperators and Defectors, $\tilde{\alpha}$ and $\tilde{\beta}$, taking into account the signaling and error reduction:
\begin{equation}
	\tilde{\alpha} = 1 - r_c \left( 1- \alpha \right)
\end{equation}
\begin{equation}
	\tilde{\beta} = 1 - r_c \left( 1- \beta \right)
\end{equation}

We'll start with calculation of relative frequencies of various types of encounters in singles allowed approach. For the `singles allowed' situation we have for encounters ending in mutually agreed match:
\begin{equation}
	F_{CC}^M = x^2 \tilde{\alpha}^2,
\end{equation}
\begin{equation}
	F_{CD}^M =F_{DC}^M = x (1-x)(1- \tilde{\beta}),
\end{equation}
\begin{equation}
	F_{DD}^M = (1-x)^2, 
\end{equation}
where $F_{XY}^M$ denotes frequency of encounter initiated  by agent type  $X$, with partner of type $Y$ and resulting in mutual consent to a match. Similarly, for encounters ending without mutual consent we have:
\begin{equation}
	F_{CC}^S = x^2 (1- \tilde{\alpha}^2),
\end{equation}
\begin{equation}
	F_{CD}^S =F_{DC}^S = x (1-x) \tilde{\beta},
\end{equation}
\begin{equation}
	F_{DD}^M = 0.
\end{equation}

In the alternative `must match'  model agents are not allowed to stay single (which may correspond to situations where a single agent would not produce any outcome).  As only Cooperators are discriminating in their choice of acceptable partners, it would seem that the change would apply only to them. In fact, the `must match' model influences matching frequencies for both Cooperators and Defectors.

To ensure that there are no unmatched agents we introduce the following scenario: if, during the initial phase of the encounter, either of the agents does not consent to a match, the initiator looks around for another partner. In principle such search might be repeated forever. Within our model, mathematically the process corresponds to fast converging geometrical series. 

Let's consider first encounters initiated by a Cooperator. Frequencies of possible outcomes are given by:

\begin{center}
\begin{tabular}{cccc}
	Initiator  &  Partner & Result  & Frequency \\ \hline
	C  &  C  &  Match  &  $x \tilde{\alpha}^2$ \\
	C  &  D  &  Match  &  $(1-x)(1-\tilde{\beta})$ \\
	C  &  C  &  No Match & $x (1- \tilde{\alpha}^2)$ \\ 
	C  &  D  &  No Match & $(1-x)\tilde{\beta}$ \\ \hline
\end{tabular}
\end{center}
Thus, the combined probability that the encounter would lead to no match situation (and thus to further search) is 
\begin{equation}
	Q_C = 1- \left[ x \tilde{\alpha}^2 + (1-x)(1-\tilde{\beta}) \right] < 1.
\end{equation}
We repeat the search process until mutual consent is ensured, which  corresponds to summing infinite geometric series. For example, probability that $C$ matches with $C$ is;
\begin{equation}
	F_{CC}^M = x \tilde{\alpha}^2 + Q_C x \tilde{\alpha}^2 + Q_C^2 x \tilde{\alpha}^2 + \ldots =
	\frac{x \tilde{\alpha}^2}{1-Q_C}.
\end{equation}
Similar analysis may be performed for situations where the initiator is a Defector:
\begin{center}
\begin{tabular}{cccc}
	Initiator  &  Partner & Result  & Frequency \\ \hline
	D  &  C  &  Match  &  $x (1-\tilde{\beta})$ \\
	D  &  C  &  No Match & $x \tilde{\beta}$ \\ 
	D  &  D  &  Always Match & $(1-x)$ \\ \hline
\end{tabular}
\end{center} 

Here, corresponding $Q_D=x \tilde{\beta} < 1 $. Finally we have full set of expressions for frequencies of encounter results:
\begin{eqnarray}
	F_{CC}^M & = & \frac{x \tilde{\alpha}^2}{x \tilde{\alpha}^2 + (1-x)(1-\tilde{\beta})} \\
	F_{CD}^M & = & \frac{x (1- x) (1- \tilde{\beta})}{x \tilde{\alpha}^2 + (1-x)(1-\tilde{\beta})} \\
	F_{DC}^M & = & \frac{x (1- x) (1- \tilde{\beta})}{1 - x \tilde{\beta}} \\
	F_{DD}^M & = & \frac{(1- x)^2 }{1 - x \tilde{\beta}}
\end{eqnarray}

\subsection{Costs}
To be able to model the evolution of our system we need to introduce the costs associated with signaling and error reduction, as well as the traditional payoffs of the matching PD games. As we try to keep our model  natural, we propose a very simple linear form of cost of signaling. For Cooperators we propose
\begin{equation}
	S_C(s(i)) = \sigma s(i),
\end{equation}
where $\sigma$ is a numerical constant (one of system parameters).  The Cooperators benefit from being recognized, but perfect recognizability ($s(i)\approx 1$) is costly. On the other hand, the Defectors, whose aim is rather to deceive than to inform, would strive to reduce $s(i)$. In our approach such deception tactics should be more expensive that normal signaling and increase with $1-s(i) \rightarrow 0 $. We propose
\begin{equation}
	S_D(s(i)) = f \sigma (1- s(i)),
\end{equation}
where the additional factor $f\geq 1$ reflects the relative difficulty of deception over straight signaling.  Grafen showed how an honest signaling system is stabilized through costly signaling: cost stabilizes the system when the cost of lying is greater than any
benefit associated with doing so \cite{grafen90-1}. It should be noted that  here we are interested  in situations where the benefits from successful match, especially Cooperator-Cooperator and Defector-Cooperator are higher than signaling costs. The reason for this assumption is to investigate if even in those unfavorable circumstances, where the Defector strategy is in principle more profitable, the cooperative strategy could emerge from nonrandom choice of partners.

To estimate the costs of error reduction we observe first that there should be no costs if $r(i) =1$ (i.e. when there is no error reduction). Moreover, very small values of $r(i)$, which allow the agent to perfectly recognize other agents, regardless of their true or deceptive signaling should be very expensive. Thus we propose cost function for error recognition to have the form
\begin{equation}
	R_C(r(i)) = \rho \left( \frac{1}{r(i)} -1 \right),
\end{equation}
with $0 < r(i) \leq 1$. The same form $R_C$ is used for both Cooperators and Defectors. It should be noted that intuitively, the error reduction is quite unimportant for Defectors, as they agree to all proposals. Thus, while `investing' in small but costly $r(i)$ should be important for Cooperators, the investment is fruitless for Defectors. They would rather benefit from minimization of costs and keep $r(i)$ close to 1.

\subsection{Payoffs and population dynamics}

Frequencies of various types of encounters, payoffs from the encounters and costs of signaling and error reduction allow to calculate the general payoffs of the Cooperators and Defectors. The difference between the payoffs determines then the population dynamics, leading to increase of the population of  the agents with higher payoff.

For the match results we use the traditional notation:
\begin{eqnarray}
	T > 0 & & \text{Payoff for Defector in match with Cooperator} \\
	R > 0  (R<T)  & & \text{Payoff for Cooperator in match with Cooperator} \\
	S = R+P-T <0 & & \text{Payoff for Cooperator in match with Defector} \\
	P  & & \text{Payoff for Defector in match with Defector} \\
	U  & & \text{Payoff for any agent staying single (if it is allowed in the model)} 
\end{eqnarray}

We further assume that $P=0$ and $U=0$. The assumption is made for simplicity and it is not generally necessary to have $P=U$ \cite{bergstrom03-3}. However it is quite natural that the payoff for each of the two Defectors working `together but each by himself' in a match should be similar to payoff of a single agent --- if acting alone is possible.

The average payoffs for Cooperators and Defectors, assuming that the agent initiates $k$ encounters in each iteration, are:
\begin{eqnarray}
	{\cal P}_C & = & \frac{2 k R F_{CC}^M + k S F_{CD}^M + k S F_{DC}^M}{x} -2 k S_C(\alpha) - 2 k R_C(r_c) \\
	{\cal P}_D & = & \frac{k T (F_{CD}^M + F_{DC}^M )}{1-x} -2 k S_D(\beta) - 2 k R_C(r_d) 
\end{eqnarray}

It is worth noting that we have assumed that the signaling and error reduction costs are paid at each \textbf{encounter} while the PD games payoffs are paid per each \textbf{match} (with $P=U=0$).

\subsection{Results of the analytical model}

The model described in our work has relatively large number of parameters. These are: payoffs from the PD game ($T, R, S, P, U$), parameters governing the costs ($\sigma, \rho, f $), initial fraction of Cooperators $x_0$, and the values of average $s(i)$ and $r(i)$ for Cooperators and Defectors ($\alpha$, $\beta$ and $r_c, r_d$).

\begin{figure}[t]
\centering
\includegraphics[height=12cm]{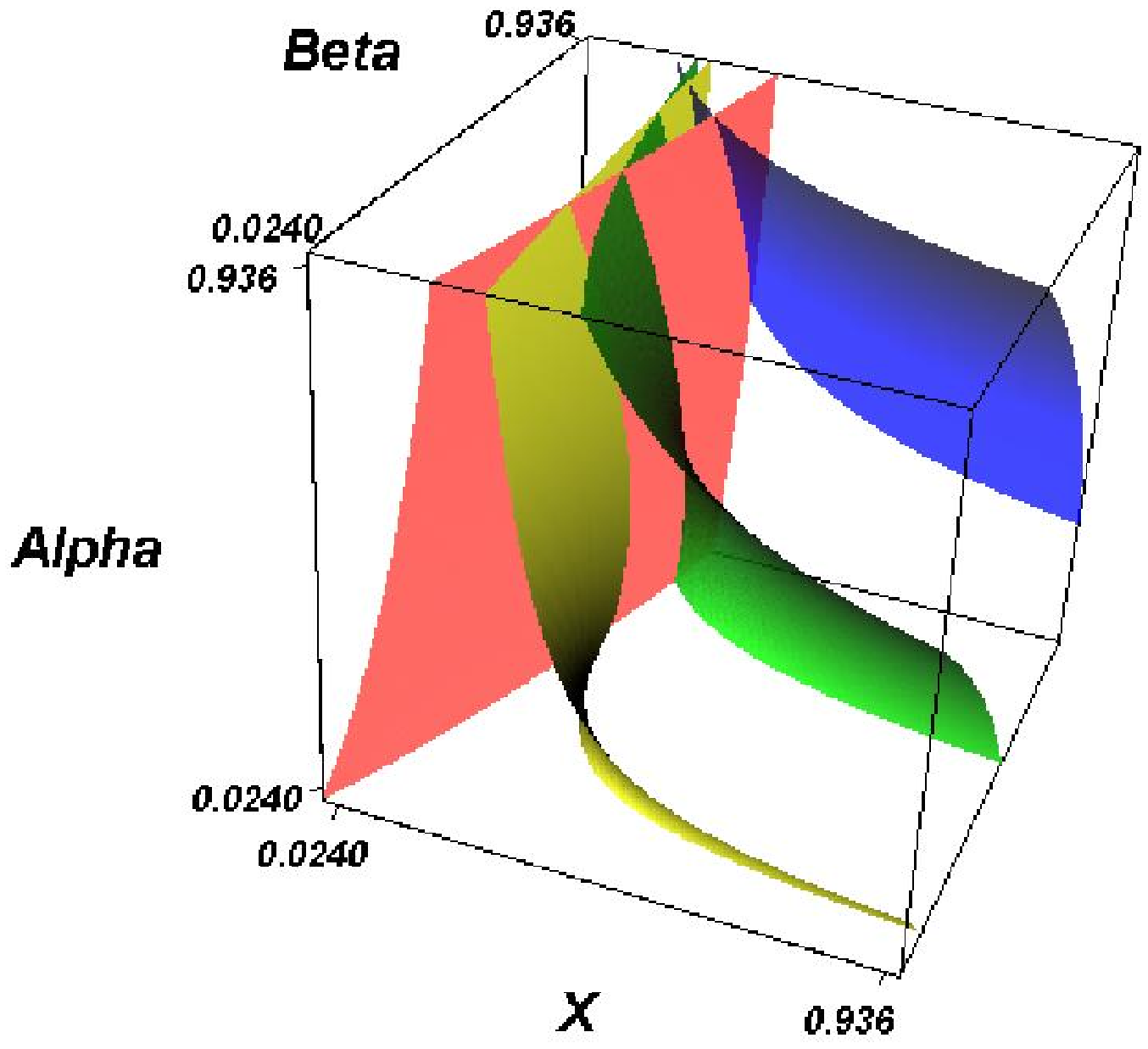}
\includegraphics[height=12cm]{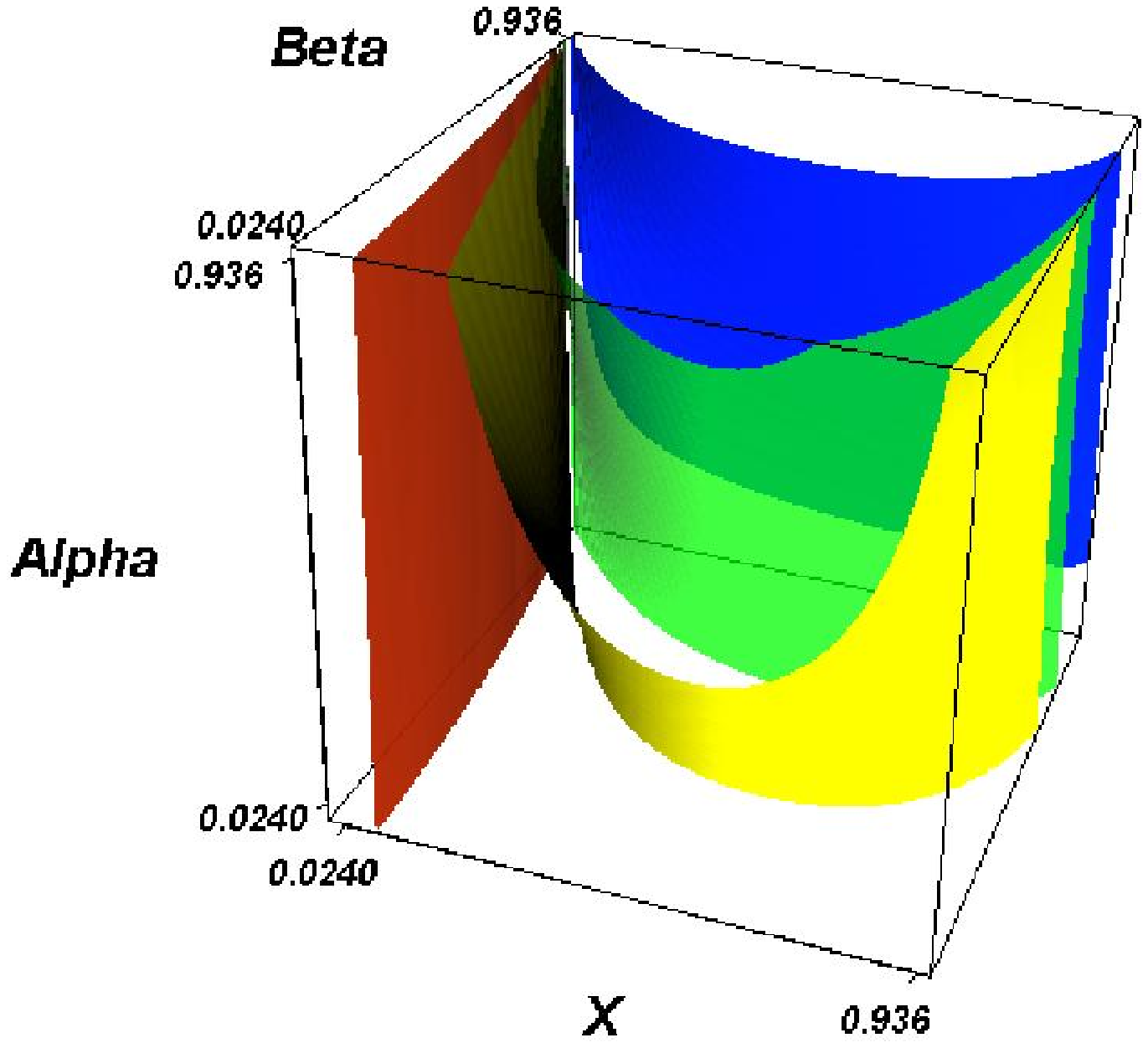}
\caption{Isosurfaces of zero payoff difference for the analytical model as functions of $\alpha$, $\beta$ and $x$, for the model allowing agents to stay single (upper figure) and for the model where they must match (lower figure).  The surfaces correspond to several values of error reduction for Cooperators, $r_c$. Red: $r_c=0.20$; yellow: $r_c=0.40$; green: $r_c=0.50$; blue: $r_c=1.00$.  Other parameters: $T=10, R=6, S=-4, P=U=0, \sigma=1.00, \delta=0.30, f=2$. \\
For the `singles OK' model Cooperator payoff is greater than the payoff of Defectors for large enough values of $x$ and $\alpha$ and $\beta$. As a result, for fixed $\alpha$, $\beta$  and $r_c$ there are either two stable outcomes of system evolution: $x=0$ and $x=1$, depending on the initial value of $x$ (if the line $\alpha,\beta,x\in[0 ,1]$ crosses the isosurface) or just  the $x=0$ result.\\
For the `must match' model, folded form of the zero isosurface results in the mixed population being possible. For values of  $\alpha$ and $\beta$ when the line $x\in[0,1]$ crosses the surface at two points $x_{\text{min}}(\alpha,\beta)$ and $x_{\text{max}}(\alpha,\beta)$ the evolution leads to $x=0$ if starting value is smaller than $x_{\text{min}}(\alpha,\beta)$, and to $x_{\text{max}}(\alpha,\beta)$ if starting value is greater than $x_{\text{min}}(\alpha,\beta)$. The situation changes for small enough values of $r_c$, where the surface is no longer folded, and the two stable outcomes are again $x=0$ and $x=1$.
\label{3d}}
\end{figure}

Figure~\ref{3d} presents examples of solutions of equation $\delta = {\cal P}_C - {\cal P}_D =0 $ for a given set of payoff parameters, as functions of $\alpha$, $\beta$ and $x$. Several surfaces, corresponding to different values of error reduction for the Cooperators ($r_c$) are shown. The surfaces correspond to points at which $\delta = 0$ in figures presented in \cite{bergstrom03-1,sobkowicz-ass}. Regions of $\delta > 0$ bound by the shown surfaces lead to increase of Cooperator population (which means increase of $x$) and vice versa, regions of $\delta<0$ favor increase of Defector population. In both models, the $\delta<0$ region starts from the $x=0$ boundary. This means that we expect the $x=0$ (Defector society) to be stable solution of the evolutionary process, if the starting conditions lie close enough to $x=0$. If  the $\delta >0$ region stretches right to the $x=1$ limit then the other stable solution is the Cooperator society. On the other hand, if the surface $\delta=0$ is folded as in the `must match' model we expect that there might be a stable mixed population solution. Keeping $\alpha, \beta, r_c$ fixed allows us to predict the final values of population composition based on initial composition. However, if we allow the parameters to change the population path might wander through the available space and analytical prediction is not possible.

\section{Computer simulations}
Computer simulations allow the study of the behavior of the society and its characteristics in a truly multidimensional parameter space. The simulations use the same costs and benefits model as the analytical formulation. We used a system with 1000 agents, each of those initiates $k$ encounters in each simulation iteration. Partial payoffs and costs from each encounter are summed up. After full cycle of encounters the agents with high total payoffs are allowed to breed, while the agents with low payoffs are eliminated from the population. Two  breeding mechanisms were used. In the first, a certain number of  `worst performers' simply took on the characteristics of the `best performers'. We call this approach `deterministic breeding'. Alternative approach of `probabilistic breeding' in which each agent compared her payoff to another randomly chosen agent, and the worse of the pair copied the characteristics of the better (this approach is sometimes called `learning model' \cite{riolo01-1}). In many cases results of the two breeding models were similar, but there were cases where significant qualitative differences ensued. This was true especially, if the number of breeding agents $N_{\text{breed}}$ was small --- in the learning model all agents can participate in breeding.  In addition to breeding dependent on agent's payoff we have introduced small fraction of  mutations: $N_{\text{mutant}}$ agents would assume randomly chosen values of $t(i)$, $s(i)$ and $r(i)$.

As for the analytic model, the results of the simulations depended on the conditions of matching (`must match' or `singles allowed') and on the input parameters. Simulations with fixed values of $s_c(i)=\alpha $, $s_d(i)=\beta $, $r_c(i)$ and $r_d(i)$ (that is simulations where the only the type $t(i)$ of an agent was changeable) produced results  in full agreement with predictions of the analytic model, with the same stable population compositions. On the other hand, simulations in which more agent characteristics ($t(i), s(i), r(i)$) were allowed to change freely have led to surprisingly rich range of outcomes. 

\begin{figure}[t]
\centering
\includegraphics[height=22cm]{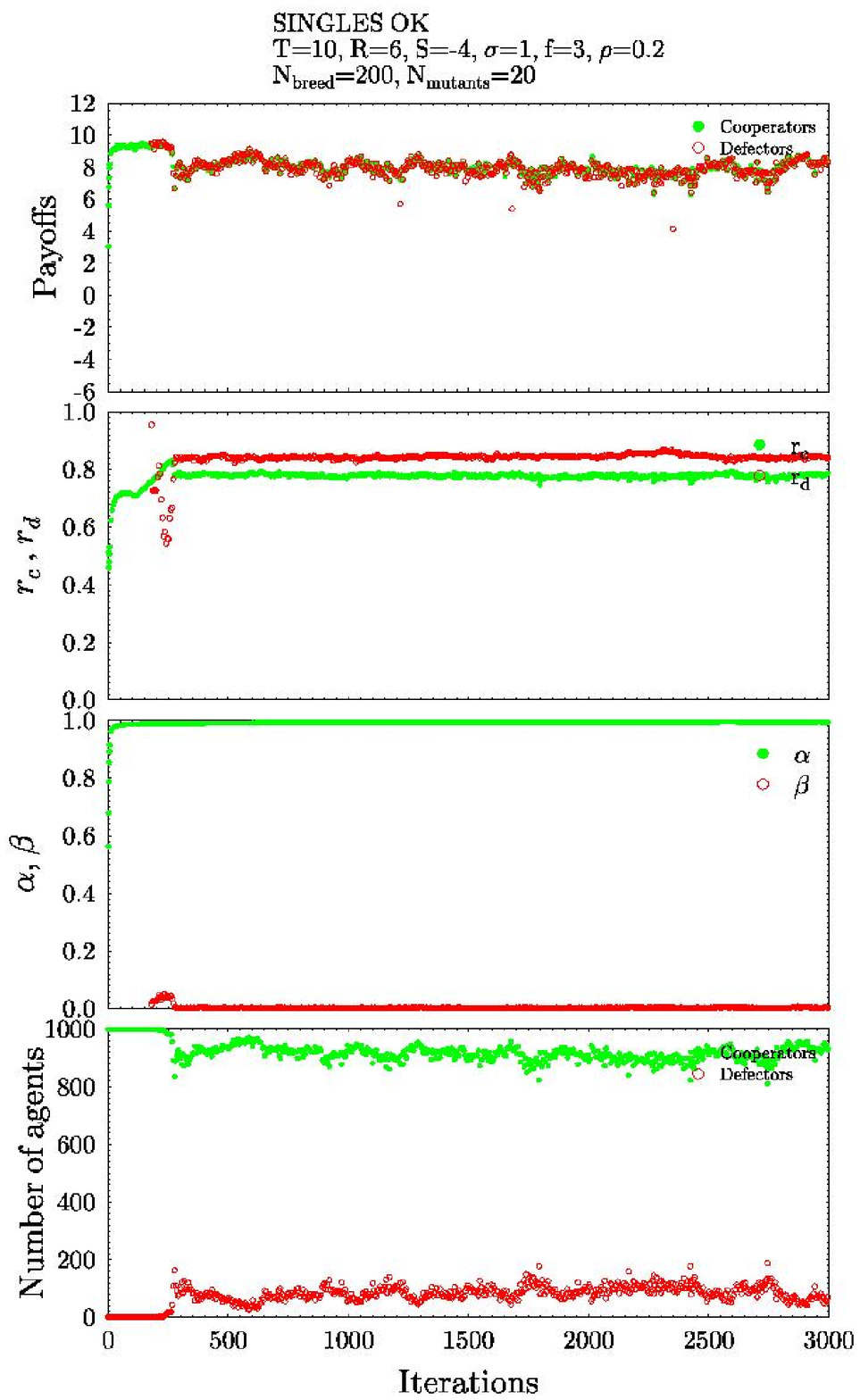}
\caption{Results of simulation for `singles allowed' model. Green points: Cooperators, red points: Defectors. Panels show evolution of number of agents of given type, average values of $s(i)$ and $r(i)$, and average payoffs. \label{sing-simple}}
\end{figure}

For some sets of simulation parameters the behavior of the system was very simple, for example either the Cooperators or Defectors quickly dominated the system, and the occasional mutants of the other type were strongly disadvantaged and eliminated. On other occasions, relatively stable mix of Cooperators and Defectors develops, fluctuating a little around a value that depends on the cost parameters. Such situation  for the `singles allowed' model is presented in Figure~\ref{sing-simple}. The four panels show, from bottom, number of Cooperators and Defectors, average values of signal strength for the two groups ($\alpha, \beta$), average values of error reduction ($r_c, r_d$) and average payoffs. Starting conditions used high value of Cooperator number (920 out of 1000) and random distribution of signals and error reduction. The simulation used deterministic breeding with top 200 performers allowed to breed at the expense of the worst 200. Twenty agents were mutated at each iteration.

The conditions necessary for such `simple' behavior are usually those that allow one type of agents clearly recognizable advantages, for example large value of $f$ ---  which makes camouflage of Defector costly, or low value of error reduction cost $\rho$ lead to Cooperator dominance. On the other hand high cost of error reduction or high value of $T$ over $R$ leads to Defector success. For some sets of input parameters the simulations result, however, in a much more complicated behavior, which can be interpreted, although not predicted in detail.

\begin{figure}[t]
\centering
\includegraphics[height=22cm]{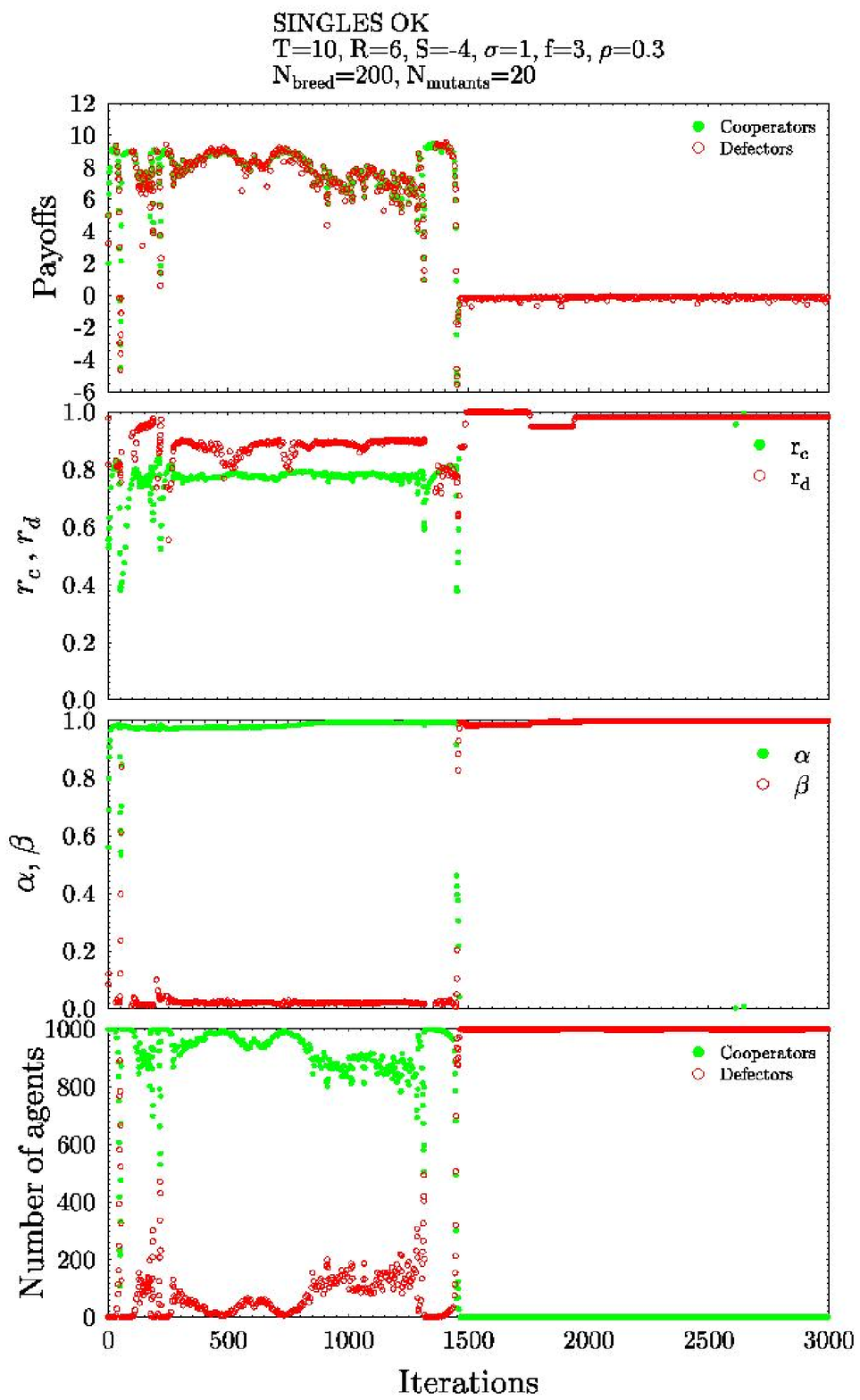}
\caption{Results of simulation for `singles allowed' model. Green points: Cooperators, red points: Defectors. Panels show evolution of number of agents of given type, average values of $s(i)$ and $r(i)$, and average payoffs. Further discussion in text.\label{sing-rich}}
\end{figure}

A good example of such behavior is provided in Figure~\ref{sing-rich}, showing results of a simulation for the `singles allowed' model.  The only difference from the input parameters of Figure~\ref{sing-simple} is the value of $\rho=0.3$ instead of 0.2, which seems at first glance a minor change indeed. Yet the system behavior is strikingly different.

The simulation results might be broken into several phases. Initially Cooperators dominated the system. There are periods of total dominance $N_C \approx 1000$, where only ephemeral mutants were Defectors. At other periods, the dominance although not absolute, was still in the range of 90\% of the population. In such situation there was significant evolutionary \textbf{internal} competition between Cooperators. This led to a race to minimize the costs, especially error reduction costs (faculty quite unnecessary in a fully Cooperator society), and thus Cooperators with higher $r_c$ (and thus lower ability to detect Defectors) were preferred. This resulted in the observed growth of $r_c$. At the same time, the Defectors were selected in a way that very strongly preferred successful cheaters ($\beta \approx 0$). This created sitation where well camoufleged Defectors could invade the Cooperator population. The periods of total Cooperator dominance have proven to be especially prone to being invaded. Several such invasions (around iteration 50, 180, 215 and 1315) were contained and reversed, mainly because the surviving Cooperators had low enough $r_c$ and were able to recognize and isolate the Defectors. However, one of the invasions (around iteration 1450) occured at the time where internal competition among Cooperators has pushed $r_c$ to very high level. Rapidly diminishing number of Cooperators had not enough `genetic variety' and in the course of just 14 iterations Defectors totally dominated the system. Soon afterwards, internal competition this time among Defectors has led to $r_d \rightarrow 1$ and $\beta \rightarrow 1$ which minimized the costs. Although mutation still produced occassional Cooperator, in the purely Defector society they were not able to combine the ability to detect Defectors and cooperate among themselves --- and the Defector dominance has proven to be stable. The complicated behavior presented here was observed in quite a few simulations. While the general picture was recognizable, particular aspects, such as the number of Cooperator dominated periods, or time of the final successful `Defector invasion' differed for different simulation parameters or even for different random number generator seeds.

\begin{figure}[t]
\centering
\includegraphics[height=22cm]{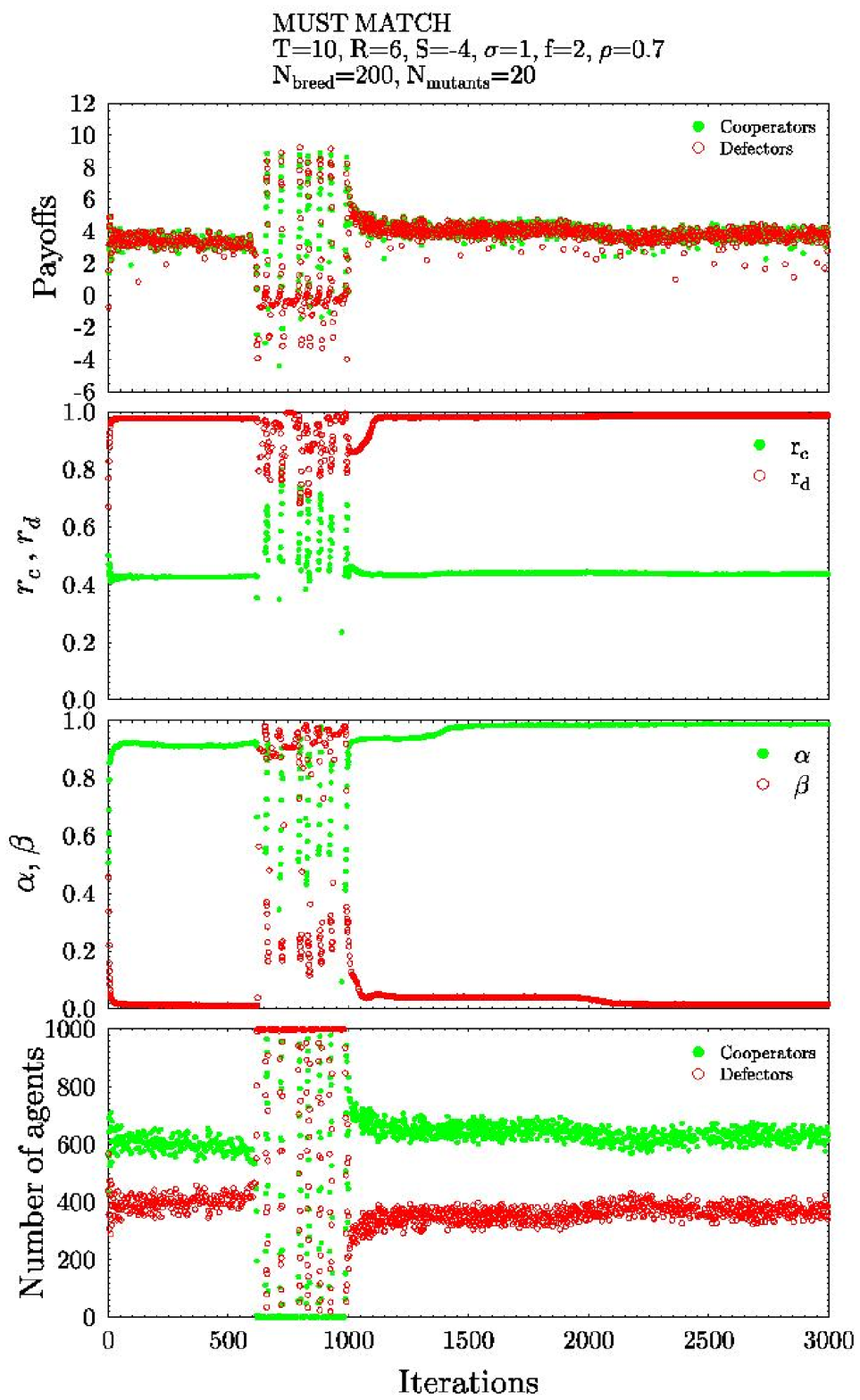}
\caption{Results of simulation for `must match' model. Green points: Cooperators, red points: Defectors. Panels show evolution of number of agents of given type, average values of $s(i)$ and $r(i)$, and average payoffs. Further discussion in text.\label{match-rich}}
\end{figure}

The `must match' model has shown comparably rich behavior. In addition to simple results of pure or mixed (but stable) societies we have observed simulations with very dramatic changes in population composition. An example is provided in Figure~\ref{match-rich}. Initially looking as a rather boring $x\approx 0.6$ system, it has entered wildly oscillating  `revolution' during which within tens of iterations society composition changed from Cooperator to Defector dominance and vice versa. The onset of these oscillations might be traced to random appearance of some Defectors with very low value of $s(i)$, who were able to successfully cheat the Cooperators. The rapid cycles during the revolution resulted from internal competition among the dominant agents. For several iterations the system was totally dominated by Defectors, during these periods the Defectors gradually decreased their ability to pose as Cooperators (increasing $\beta$) which opened the way for equally brief attempts by mutant Cooperators to gain strength. The Cooperator incursions lasted also several iterations. The end of the `revolution' phase is equally rapid and mysterious. One of the Cooperator invasions resulted in a group with sufficiently low $r_c$ value --- enabling them to successfully recognize Defectors. Finally new equilibrium formed, with populations of Cooperators and Defectors close to the pre-revolutionary times, but with values of $\alpha, \beta$ and $r_c$ that  blocked the future revolutions. Even though we have tested our system through several thousands of iterations the `revolutionary' phase was not repeated. It should be noted here that using slightly different values of input parameters (including the random number generator seed) resulted in similar behavior, although happening at different simulation timeframe.

\begin{figure}[t]
\centering
\includegraphics[height=22cm]{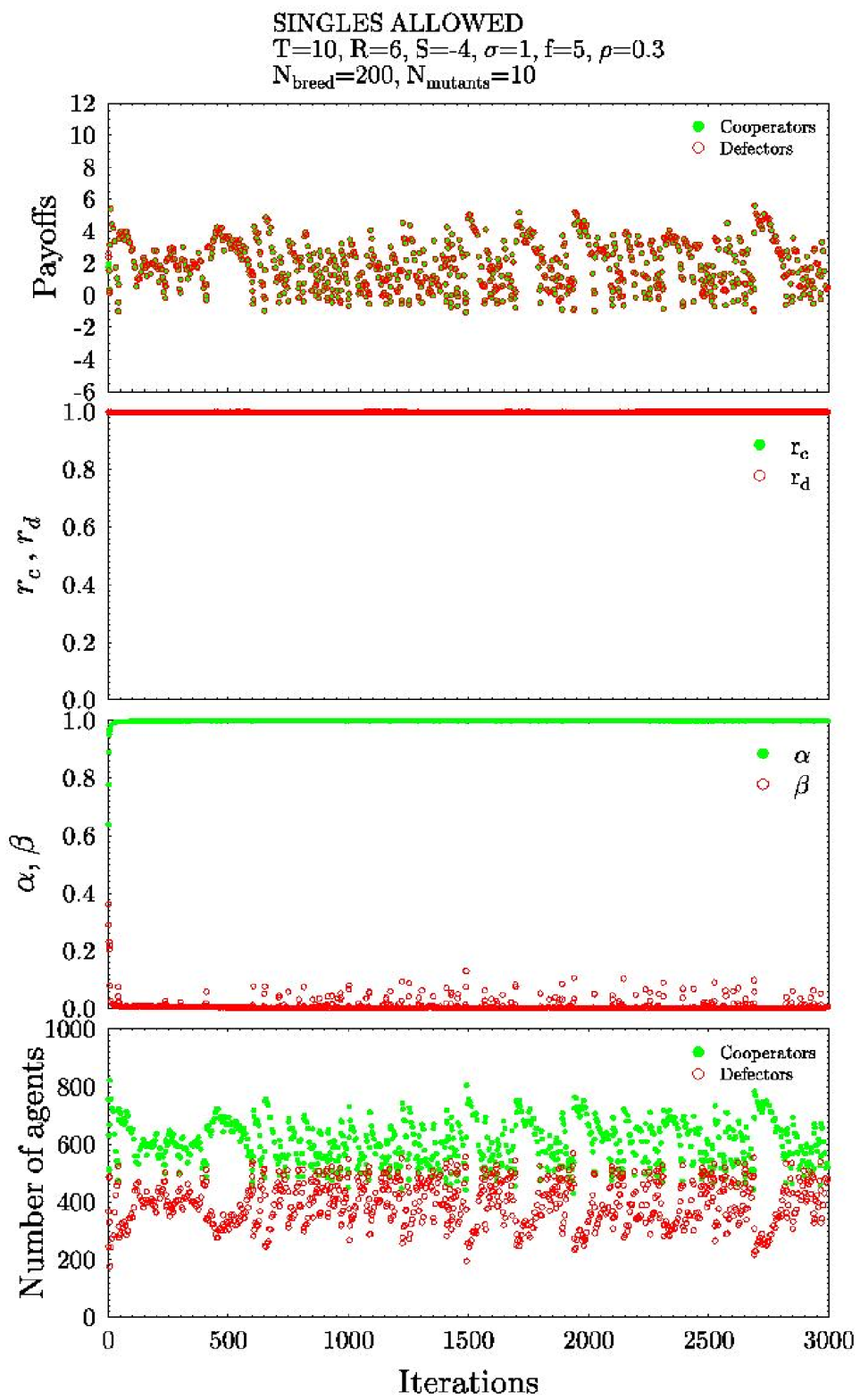}
\caption{Example of results of simulation for `singles allowed' model with no error reduction. Green points: Cooperators, red points: Defectors. Panels show evolution of number of agents of given type, average values of $s(i)$ and $r(i)$, and average payoffs. Further discussion in text.\label{fixrc-rich}}
\end{figure}

So far we have presented examples of \textit{complex behavior} for a fully free system, in which all $t(i), s(i)$ and $r(i)$ could change within the available range. Some of the phenomena we have presented were clearly linked to error reduction by Cooperators. It is interesting whether any comparably complicated activity could be observed in a system where there is no error reduction. To test this we have simulated the system where $r(i) \equiv 1$ for all agents. The only factors determining agent recognition was the strength of the signaling, whether true or misleading. This situation corresponds directly to that presented in \cite{bergstrom03-1,sobkowicz-ass}. Generally, results of the simulations for such system were  simpler, leading to more often to stable pure or mixed societies. However, even  such simplified model could lead to quite complex evolution of the system. Figure~\ref{fixrc-rich} presents repeating, aperiodic cycles of changes in the population composition for the `singles allowed' model. This shows that simple conclusions from one dimensional analyses of earlier works might miss some interesting aspects of the influence of the assortativity of matching process to the emergence of cooperation. The evolving system is capable of more than just converging to fixed points in population composition $x$.

One obvious property of the model is that in all situations where Cooperators dominate the system, the average payoff per agent is higher. Achieving this \textbf{group benefit} within biological systems where selection acts on individual level (or even on the sub-individual gene level) is difficult, and happens mostly through  mechanisms other than assortative matching (such as kin selection or reciprocal altruism). It is, however, possible for socially constructed systems (such as trading environment) to impose the conditions that would increase the chances of cooperativeness to prosper. For example the universal standards of accounting and audit  are a way of \textbf{imposing} costly signaling aimed at keeping out the Defectors. The companies can recognize prospective business partners `strategy' in PD game like scenario by looking into public financial records. This would correspond, in our model to putting $f$ very high (as the costs of successfully camouflaging company results are prohibitively high). In such situations Cooperators have much better chances to dominate the system in a stable way.

\subsection{Simulation time vs. real time: the role of metastable states}

An interesting aspect of any computer simulation is the artificiality of the `iteration time'. depending on input parameters, such as the number of breeding agents  or  the amount of encounters within a given iteration $k$  the tempo of system evolution is changed. In most cases, though, the changes of these parameters merely speeded up or slowed down the system evolution as measured in iterations. Sometimes --- for example when the number of breeding agents is small, qualitatively different results are obtained, usually attributable to locally stable systems, for which small number of agents that are allowed to breed can not disturb the equilibrium. Yet it remains an important task to find any connection between this artificial time and real timescales of the phenomena we set out to model, such as biological evolution of human (and non-human) societies, or evolution of cooperative behavior in various activities, such as economy or education. Direct mapping of `simulation events' (such as encounters and matches) to real life counterparts is in most cases very difficult, as the simulation only catches one of the many facets of the real environment. 

We would like to note  that our simulations have used relatively large number of individual encounters within each iteration (between 100 and 1000). While such number of encounters is not unimaginable for small societies or for merchant activities (especially in the electronic commerce age) the large number of iterations, corresponding to biological generations or to successful trader or company `generations' is way beyond the scope of the general stability of the whole system. Fifty human generations would span about 1000 years --- timescale during which changes in external conditions would bring make the constant payoff model of isolated society assumption totally absurd. The same reasoning applies to the trade activities. There are two conclusions from the above observation.

First, for comparatively short lived societies our model might still provide valuable insight, but instead of the final stable results the transient, metastable phases that we have noted, for example in Figure~\ref{sing-rich} might correspond to what we actually observe in life! Keeping the above mentioned example in mind, we might find high levels of cooperation form not because they are ultimately the most stable evolutionary solution, but because the system timeframe is such that the transition to Defector society did not yet take place. For such situation the \textbf{initial} dynamics of the system plays a crucial role. Our simulations, starting from random distribution of signaling strength and error reduction usually have very short initial phase, lasting a few iterations, during which the system very quickly evolves to the initially preferred state. Whether that state is really stable or just metastable becomes visible only after the internal evolution among the `winning' type of the agent strategy forces the winners to reduce costs. 

The second conclusion points out one very important example of a real system which could be modeled with our approach, and yet last long enough for thousands of generations in relatively stable environment. The system is the early human evolution, lasting indeed for many thousands of generations, where small groups of hominids were working together, and where assortativeness of matching would indeed lead to increased payoff for group members and evolutionary benefits. The fact that current human behavioral makeup shows significant fraction of cooperative strategy indicates that in this case the `model parameters' were those leading to at least some Cooperators surviving in long term. Whether in activities having direct link to genetic trait transmission and evolution (such as mate choice and childrearing investment) or in simpler and more frequent acts of group activities (hunting, managing tasks impossible for single individual) the ability to recognize a good prospective partner plays a crucial role. Of course, the memory effects (totally absent in our approach) play important part, but in reality the process of choosing the partner is sometimes based on memories of previous encounters and sometimes on `first glance' assessment (and therefore on signaling and lie detection) \cite{frank88-1}. It is possible to include the memory effect in our model (for example by drastically diminishing the error reduction costs for those agents that the initiator has matched with in the past). Such a synthesis of willful signaling and memory based recognition will be the subject of future investigation.

\end{document}